\documentclass[11pt,twoside]{article}

\usepackage{asp2006}
\usepackage{epsf}
\usepackage{psfig}
\usepackage{lscape}
\usepackage{natbib}
\usepackage{graphicx}

\newcommand       \msun        	{$M_{\odot}$}
\newcommand       \lsun      	{$L_{\odot}$}

\newcommand       \mic        	 {$\mu$m}

\newcommand      \yd      {$\widehat Y_d$}

\newcommand      \misme     {\left<m_{\rm ISM}\right>}
\newcommand      \mism     {$\left<m_{\rm ISM}\right>$}

\newcommand		\jay			{J1148+5251}
\newcommand      \ydust      {$Y_d$}
\newcommand      \yduste      {Y_d}
\newcommand 	     \mstar      {$m_{\star}$}
\newcommand 	     \mstare      {m_{\star}}

\newcommand		\mug			{$\mu_g$}
\newcommand		\muge		{\mu_g}

\begin{document}
\bibliographystyle{/Users/edwek/Library/texmf/tex/latex/misc/aastex52/aas.bst}

\title{The Cycle of Dust in the Milky Way: Clues from the High-Redshift and Local Universe}   
\author{Eli Dwek}   
\affil{Observational Cosmology Lab, Code 665, NASA Goddard Space Flight center, Greenbelt, MD 20771}    
\author{Fr\'ed\'eric Galliano}
\affil{Service d'Astrophysique, CEA/Saclay, L'Orme des Merisiers, 91191 Gif-sur-Yvette, France}
\author{Anthony Jones}
\affil{Institut d'Astrophysique Spatiale (IAS), B\^atiment 121,
Universit\'e Paris-Sud 11 and CNRS, F-91405 Orsay, France}

\begin{abstract} 
Models for the evolution of dust can be used to predict global evolutionary trends of dust abundances with metallicity and examine the relative importance of dust production and destruction mechanisms. Using these models, we show that the trend of the abundance of polycyclic aromatic hydrocarbons (PAHs) with metallicity is the result of the delayed injection of carbon dust that formed in low mass asymptotic giant branch (AGB) stars into the interstellar medium. The evolution of dust composition with time will have important consequences for determining the opacity of galaxies 
and their reradiated thermal infrared (IR) emission. 
Dust evolution models must therefore be an integral part of 
population synthesis models, providing a self-consistent link between the 
stellar and dust emission components of the spectral energy distribution (SED) of galaxies.
We also use our dust evolution models to examine the origin of dust at redshifts $> 6$, when only supernovae and their remnants could have been, respectively, their sources of production and destruction. Our results show that unless an average supernova (or its progenitor) produces between 0.1 and 1~\msun\ of dust, alternative sources will need to be invoked to account for the massive amount of dust observed at these redshifts.   
  
\end{abstract}

\section{Historical Overview}
Chemical evolution (CE) models follow the formation, destruction, abundance, and spatial and stellar distribution of elements created during the nucleosynthesis era of the Big Bang, and the different evolutionary stages in stars. Models are pitted against a host of observational test, such as the relative abundances of the various elements and their isotopes in meteorites, stars, and in the interstellar, intergalactic and intracluster media, the G-dwarf metallicity distribution, and the age-metallicity relation of the various systems [e.g. \citep{matteucci01,pagel97}]. 

CE models provide a natural framework for studying the evolution of dust since the abundance of the elements locked up in dust must be constrained by the availability of refractory elements in the interstellar medium (ISM). CE models need then to be generalized to include processes unique to the evolution of dust: the condensation efficiency of refractory elements in stellar ejecta, the destruction of grains in the ISM by expanding supernova remnants (SNRs), and the growth and coagulation of grains in clouds \cite{dwek98}.

The first dust evolution models \citep{dwek79,dwek80b} (hereafter DS) addressed the origin of the elemental depletion pattern, which was a subject of considerable debate. \cite{field74} showed that the depletion pattern correlated with condensation temperature, suggesting that it reflects the condensation efficiency of the elements in their respective sources. Such causal correlation requires that dust undergoes very little interstellar processing that can alter the depletion pattern. An equally good correlation exists between the magnitude of the elements' depletion and their first ionization potential \citep{snow75}, suggesting that the depletion pattern may instead be governed by accretion processes in molecular clouds. Pointing out the intrinsic physical correlation between the condensation temperature, the first ionization potential, and the threshold for grain destruction by sputtering, DS suggested that the depletion pattern could reflect the destruction efficiency of dust in the ISM. It is currently clear that all three processes play some role in establishing the elemental depletion pattern, since it depends on the density of the medium in which it is observed \citep{savage96}. Globally, their relative importance depends on the prevalence of and the cycling times between the different phases (hot, neutral, and molecular) of the ISM. 

A more detailed review of dust evolution models is given by \cite{dwek98}. Since then, several models have been constructed to follow the evolution of dust in dwarf galaxies \citep{lisenfeld98}, Damped Ly$\alpha$ systems \citep{kasimova03}, the Milky Way galaxy \citep{zhukovska08}, and the origin of isotopic anomalies in meteorites \citep{clayton04,zinner06a,zinner06b}. 

Early dust evolution models adopted the instantaneous recycling approximation, which assumes that all the elements and dust are promptly injected back into the ISM following the formation of their nascent stars. In contrast \cite{dwek98} and \cite{morgan03} constructed models that take the finite main sequence (MS) lifetimes of the stars into account. Silicate dust is primarily produced in supernovae (SN) that ``instantaneously'' recycle their products back to the ISM, whereas carbon dust is mainly produced in low-mass carbon-rich stars which have significantly longer MS lifetimes. Consequently these models predicted that the composition of the dust should evolve as a function of time. 

Dust evolution models contain many still uncertain parameters such as the dust yields in the various sources and the grain destruction efficiency in the ISM. However, in spite of these uncertainties we will show that they can be very successful in predicting global evolutionary trends, namely the observed correlation of the abundance of polycyclic aromatic hydrocarbon (PAH) molecules with metallicity, and in examining the origin of dust in the high-redshift universe. 

\section{Main Ingredients of Dust Evolution Models}

\subsection{The Yield of Dust in Stars}
The dust condensation efficiency, and its composition and size distribution depend very much on the environment in which it is formed. Dust can form in the quiescent outflows of AGB or Wolf-Rayet stars, or in the explosively expelled ejecta of SNe and novae. There is substantial observational evidence for the formation of dust in all these sources, however, their relative importance as sources of interstellar dust, and the composition and size distribution of the condensed dust are still uncertain, especially in SNe.   
Because of the ease of formation and stability of the CO molecule, dust sources can be divided into two categories: carbon-rich sources, in which the C/O abundance ratio is larger than 1. These sources will produce carbon dust; and oxygen-rich sources with C/O $<$ 1, which will produce silicate-type dust. Such simple arguments assume that CO formation goes to completion, which may not always be the case, as suggested by the dynamical condensation models of \cite{ferrarotti06}. 
 
\begin{figure}
    \begin{center}
\includegraphics[width=4.0in]{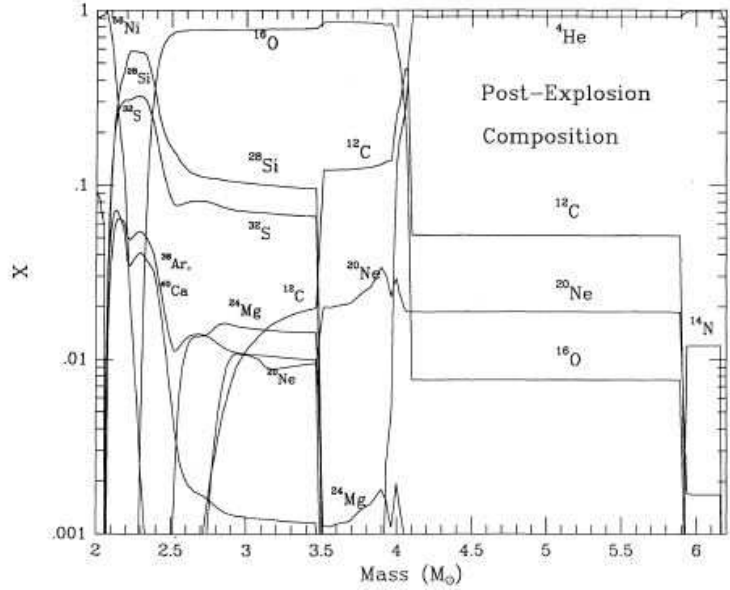} 
  \end{center} 
 \caption{The post-explosive composition of a 25~\msun\ star \citep{woosley88}. A typical SN can potentially produce 1~\msun\ of dust. }
   \label{snejecta}
\end{figure}

SN explosions mark the death of stars more massive than $\sim8$~\msun. Figure \ref{snejecta} depicts the post-explosive composition of a 25~\msun\ star \citep{woosley88}. It depicts a typical onion-skin structure in which the composition of the different layers reflect the pre- and post-explosion nuclear burning stages of the star. Globally, the ejecta has a C/O ratio $<$~1, and should in principle be only producing silicate dust. However, in spite of the mixing between the layers caused by Rayleigh-Taylor instabilities in the ejecta, this mixing is of macroscopic nature and does not occur on an atomic level. So the layers above above $\sim~4.2$~\msun, in which C/O $>$~1, will maintain this ratio and produce carbon dust, whereas the inner layers will produce silicates. If all condensible elements precipitated out of the ejecta and formed dust, the yield of dust in a typical SN would be about 1~\msun\ \citep{kozasa89,kozasa91}.

\begin{figure}
  \hspace{-0.5in}
  \begin{tabular}{cc}
\includegraphics[width=2.8in]{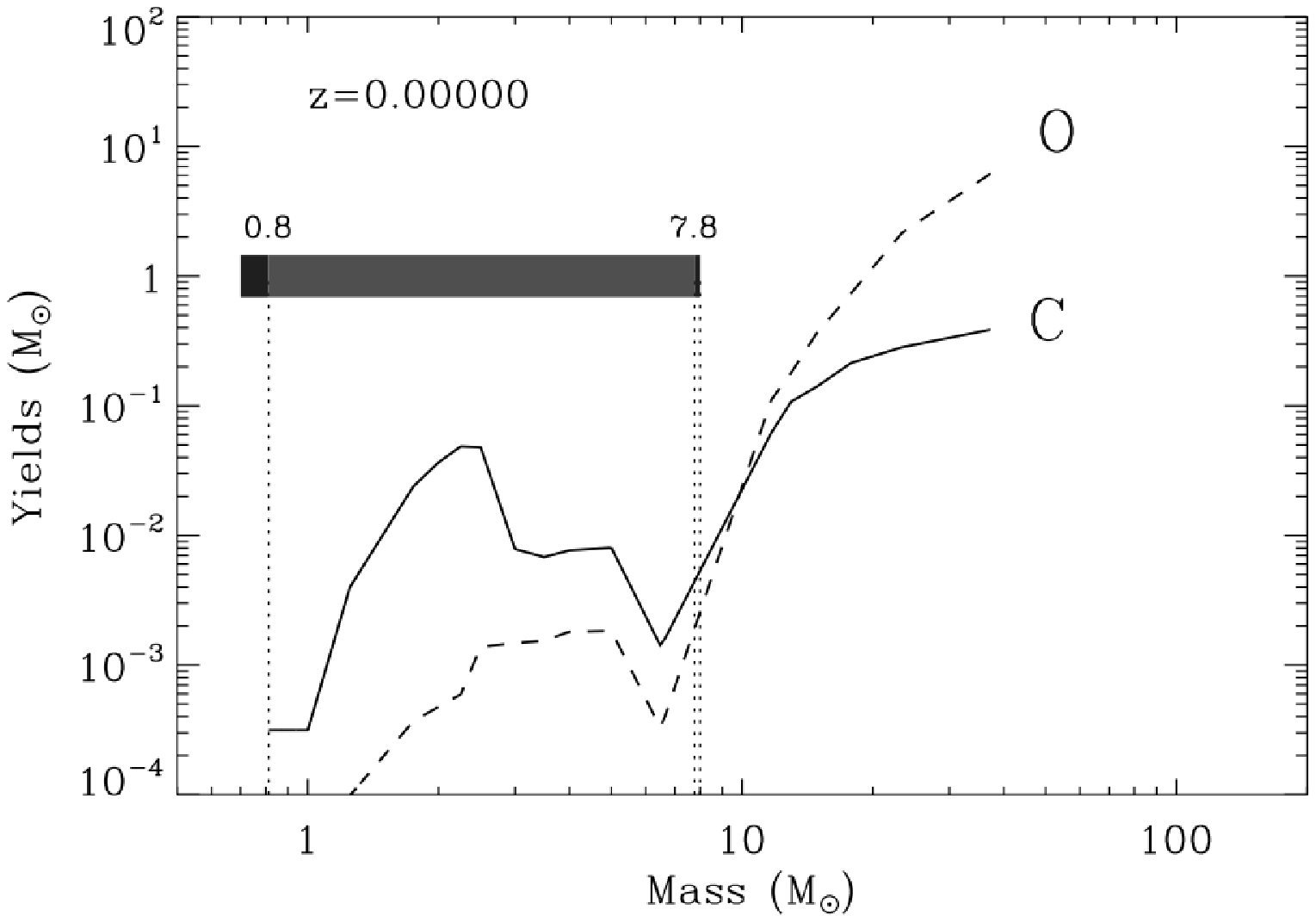}  
\includegraphics[width=2.8in]{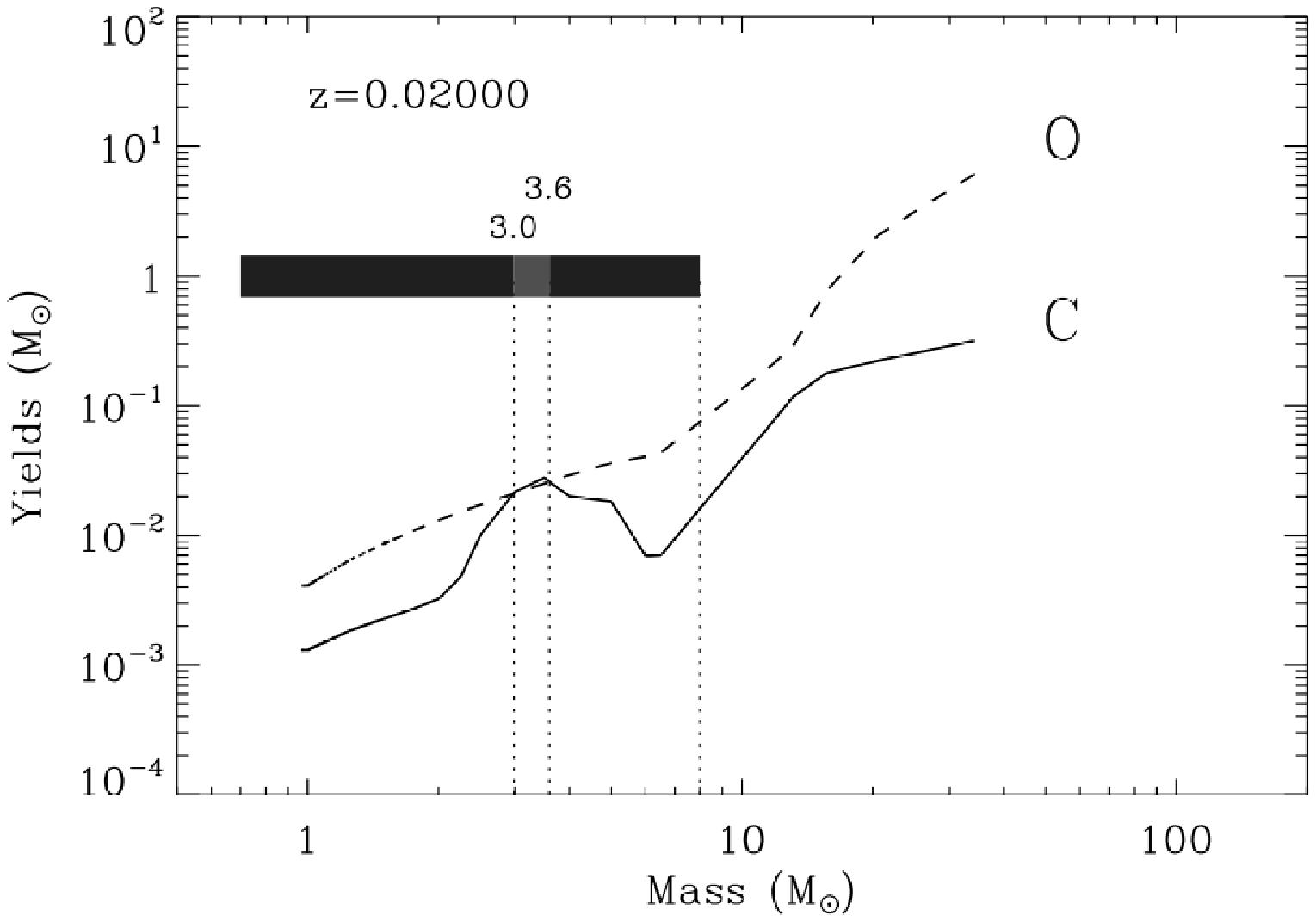}
\end{tabular}
 \caption{The C and O yields in stars. Stars with masses above $\sim~8$~\msun\ become SNe and produce both silicate and carbon dust. Stars with masses below $\sim 8$~\msun\ produce either carbon or silicate dust, depending on the C/O ratio in their ejecta. The light gray area in the horizontal bar depicts the range of stellar masses in which C/O $>$ 1.}
    \label{coyield}
\end{figure}

\begin{figure}
  \begin{tabular}{cc}
\includegraphics[width=2.5in]{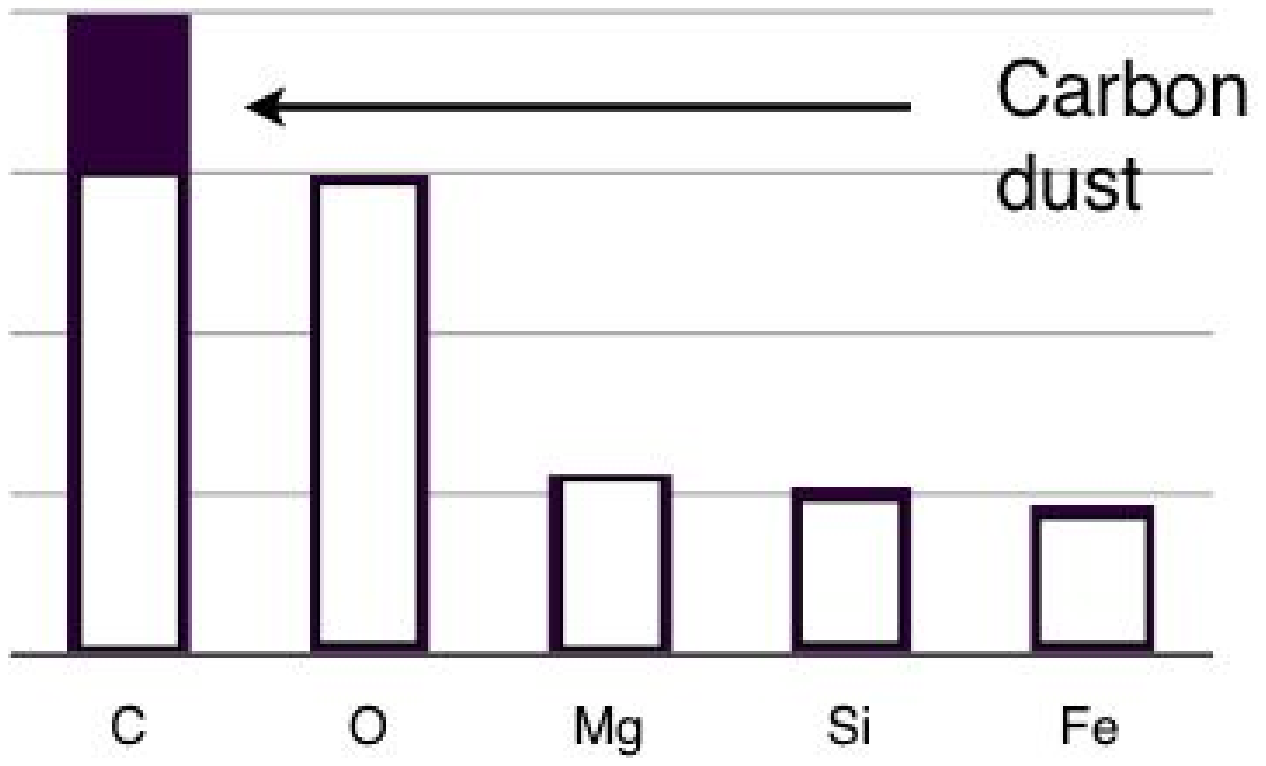}  
\includegraphics[width=2.5in]{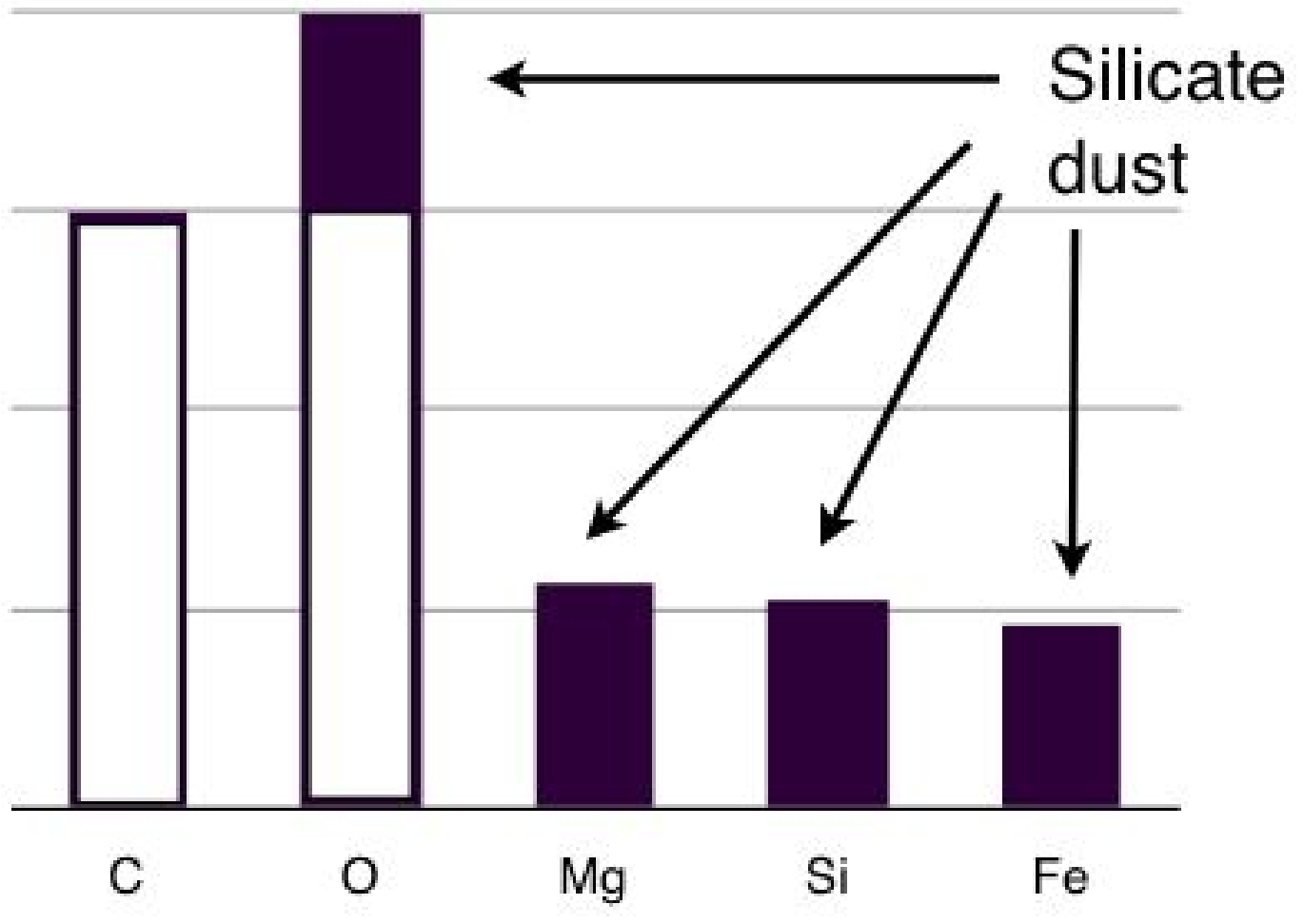}
\end{tabular}
 \caption{Qualitative depiction of the calculated yield of carbon and silicate dust in AGB stars. When $C/O > 1$ (left panel), the star produces only carbon dust. The dark shaded area depicts the number of carbon atoms that condense into dust. When $C/O <1$ (right panel) the star produces silicate dust, and the dark shaded area depicts the condensing elements.}
    \label{co}
\end{figure}

The yield and composition of dust in lower mass stars depends on the C/O ratio in their atmosphere during the AGB phase of their evolution.  
 Figure \ref{coyield} depicts the C and O yield in stars. Supernovae yields were taken from \cite{woosley95} and AGB yields were taken from \citep{karakas03a}. Stars with masses above $\sim 8$~\msun\ produce carbon and silicate dust, irrespective of the global C/O ratio in their ejecta. The mass range of carbon rich stars depends on the initial stellar metallicity. At zero metallicity (left panel) the mass range of stars producing carbon dust is between 0.8 and 7.8~\msun. This range narrows significantly to masses between 3.0 and 3.6~\msun\ at solar metallicity (right panel). Figure \ref{co} presents a qualitative depiction of the method we use to calculate the yield of carbon and silicate dust in AGB stars. A more realistic model calculating the yield of dust in AGB stars was presented by \cite{ferrarotti06} and Hoefner (2009, this conference proceedings).

\begin{figure}
  \hspace{-0.5in}
  \begin{tabular}{cc}
\includegraphics[width=2.1in]{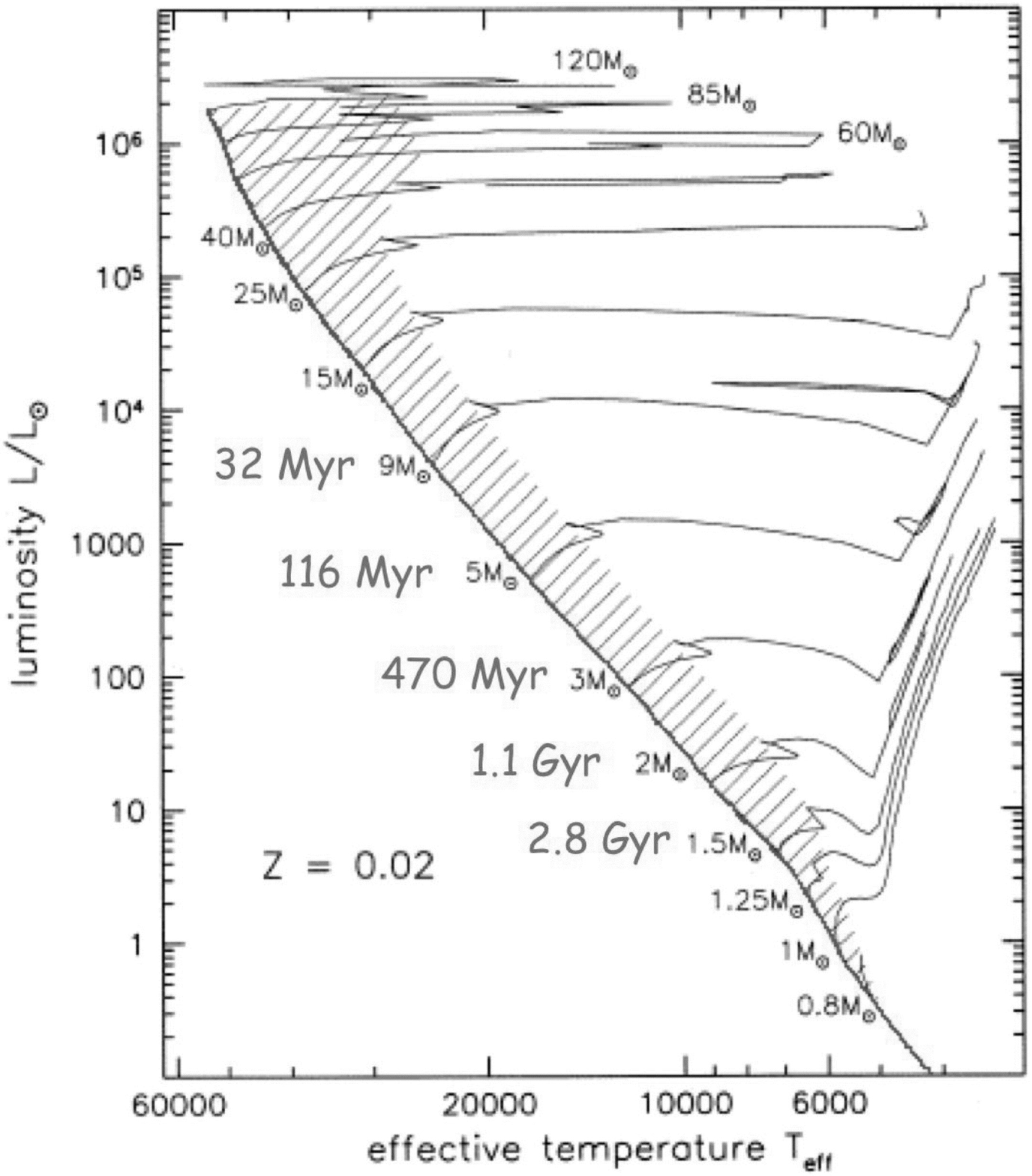} 
\includegraphics[width=3.4in]{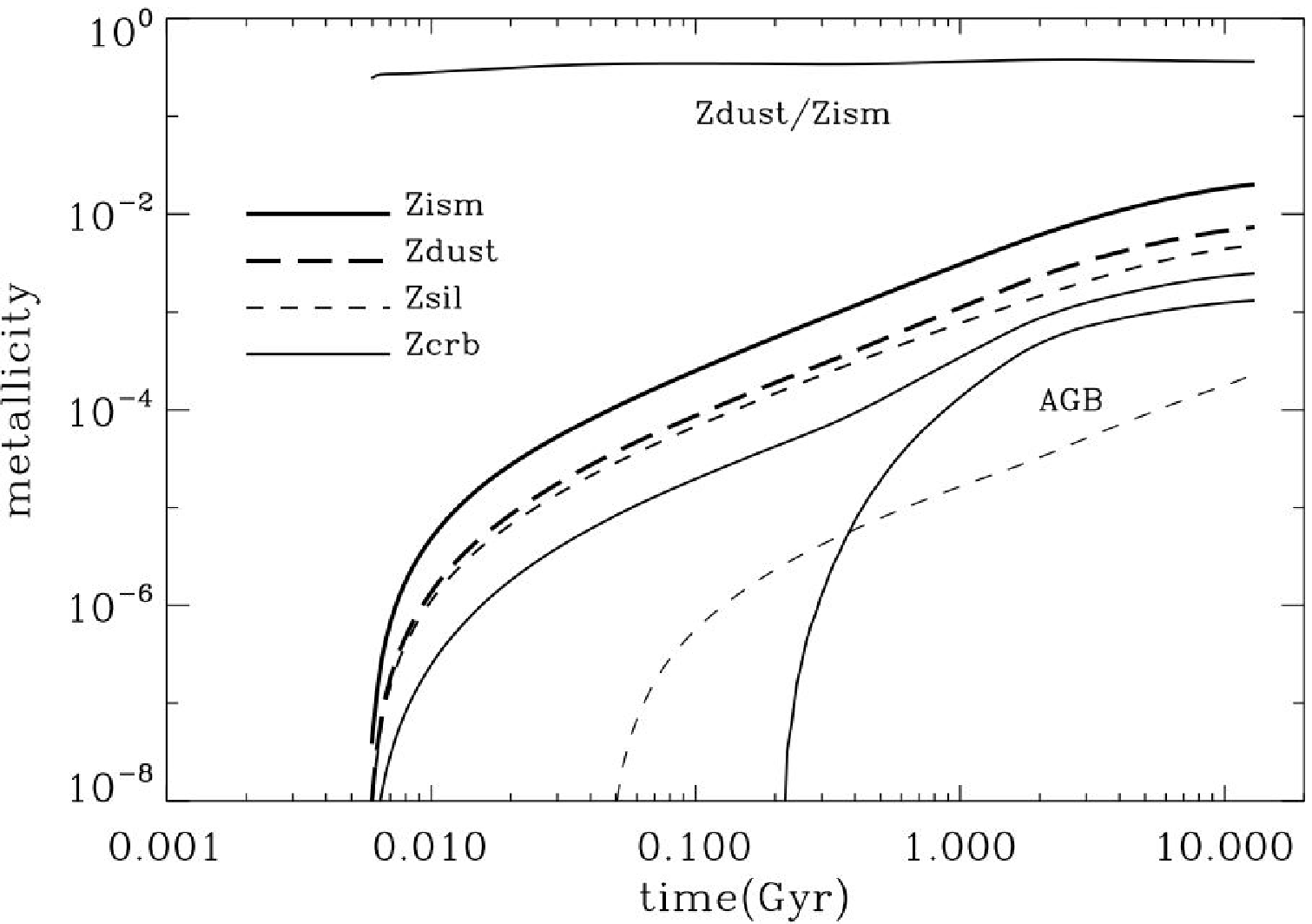}
\end{tabular}
 \caption{{\it Left panel}: The H-R diagram of stars and their main sequence lifetime. {\it Right panel}: The evolution of silicate (dashed line) and carbon (solid line) dust from SNe (bold curves) and AGB stars (light curves).}
    \label{dustvol}
\end{figure}

Figure \ref{dustvol} (left panel) depicts the stellar evolutionary tracks of stars with an initial solar metallicity. Also shown in the figure is their MS lifetime. Similar lifetimes are obtained for different initial metallicities \citep{portinari98}. At metallicity of $Z = 0$, the first carbon dust producing stars are about 8~\msun\ and will evolve of the MS about 50~Myr after their formation. At solar metallicities, the production of carbon by AGB stars will be delayed by about 500~Myr, when $\sim 4$~\msun\ stars evolve off the MS. Since most of the interstellar carbon dust is made in AGB stars, this delay in its formation can have important observational consequences. Carbon dust has a significantly higher visual opacity than silicates, so the opacity of galaxies will change with time, with young systems being more transparent than older ones.  

Figure \ref{dustvol} (right panel) depicts the different evolutionary trends of SN- and AGB-condensed dust calculated for a CE model with by exponential star formation rate characterized by a decay time of 6~Gyr, and a Salpeter initial mass function \citep{dwek98, dwek05}. 
The silicate and carbon dust yields were calculated assuming a condensation efficiency of unity in the ejecta, and grain destruction was neglected. The model therefore represents an idealized case, in which grain production is maximized, and grain destruction processes are totally ignored.  Also shown in the figure are the separate contributions of AGB stars to the abundance of silicate and carbon dust. The onset of the AGB contribution to the silicate abundance starts at $t \approx 50$~Myr, when $\sim$ 8~M$_{\odot}$ stars evolve off the main sequence, whereas AGB stars start to contribute to the carbon abundance only at $t \approx 500$~Myr, when  4~M$_{\odot}$ stars reach the AGB phase. The figure also presents the dust-to-ISM metallicity ratio, which is almost constant at a value of $\sim$ 0.36. 
 
 \section{The Lifetime of Interstellar Dust}
Following their injection into the ISM, the newly-formed dust particles are subjected to a variety of interstellar processes resulting in the exchange of elements between the solid and gaseous phases of the ISM, including: (a) thermal sputtering in high-velocity ($>$200 km~s$^{-1}$) shocks; (b) evaporation and shattering by grain-grain collisions in lower velocity shocks; and (3) accretion in dense molecular clouds.
Detailed description of the various grain destruction mechanisms and grain lifetimes in the ISM were presented by \citep{jones96, jones04}.  In addition, SN condensates can be destroyed shortly after their formation by reverse shocks that travel through the expanding ejecta (Dwek 2005, Bianchi \& Schneider 2007, Nozawa et al. 2008).
 
\begin{figure}
    \begin{center}
\includegraphics[width=4.0in]{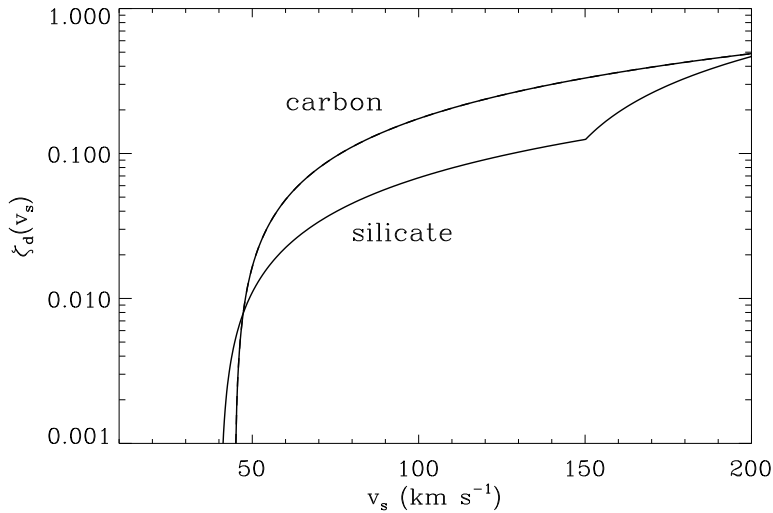} 
  \end{center} 
 \caption{The mass fraction of carbon and silicate dust that is destroyed  as a function of shock velocity (after Jones et al. 1996).  }
   \label{xsi}
\end{figure}

The most important parameter governing the evolution of the dust is its lifetime, $\tau_d$, against destruction by SNRs. In an interstellar medium with a uniform dust-to-gas mass ratio, $Z_d$, this lifetime is given by \citep{dwek80b,mckee89b}:
\begin{equation}
\tau_d = {M_d(t)\over \left<m_d\right> R_{SN}} = {M_g(t)\over \misme R_{SN}}
\label{dust_life}
\end{equation}
where $M_d$ and $M_g$ are, respectively, the total mass of dust and gas in the galaxy, $\left<m_d\right>$ is the total mass of elements that are locked up in dust and returned by a single SNR back to the gas phase either by thermal sputtering or evaporative grain-grain collisions. $R_{SN}$ is the SN rate in the galaxy, so that the product $\left<m_d\right>\, R_{SN}$ is the destruction rate of dust in the ISM. The parameter $\misme \equiv \left<m_d\right>/Z_d$ is the effective ISM mass that is completely cleared of dust by a single SNR, given by \citep{dwek07b}:
\begin{equation}
\misme = \int_{v_0}^{v_f}\ \zeta_d(v_s)\ \left|{dM\over dv_s}\right|\ dv_s
\end{equation}
where  $ \zeta_d(v_s)$ is the fraction of the mass of dust that is destroyed in an encounter with a shock wave with a velocity $v_s$, $(dM/dv_s)dv_s$ is the ISM mass that is swept up by shocks in the [$v_s,\ v_s+dv_s$] velocity range, and $v_0$ and $v_f$ are the initial and final velocities of the SNR. 
Figure \ref{xsi} depicts the mass fraction of carbon and silicate dust that is destroyed after being swept up by a shock of velocity $v_s$ as a function of shock velocity. An updated version for the carbon and PAH destruction efficiency was presented by Jones et al. (2009, this conference proceedings). For example, in the Milky Way $M_g \approx 5\times 10^9$~\msun, $R_{SN} \approx 0.03$~yr$^{-1}$, and $\misme \approx 300$~\msun, giving a dust lifetime of $\sim 6\times 10^8$~yr.

In addition, SN condensates can be destroyed shortly after their formation by reverse shocks that travel through their expanding ejecta (Dwek 2004, Nozawa et al. 2008).
 
\section{PAHs and Silicate Dust as Tracers of AGB- and SN-condensed Dust}
\begin{figure}
  \begin{center}
  \begin{tabular}{cc}
\includegraphics[width=2.2in]{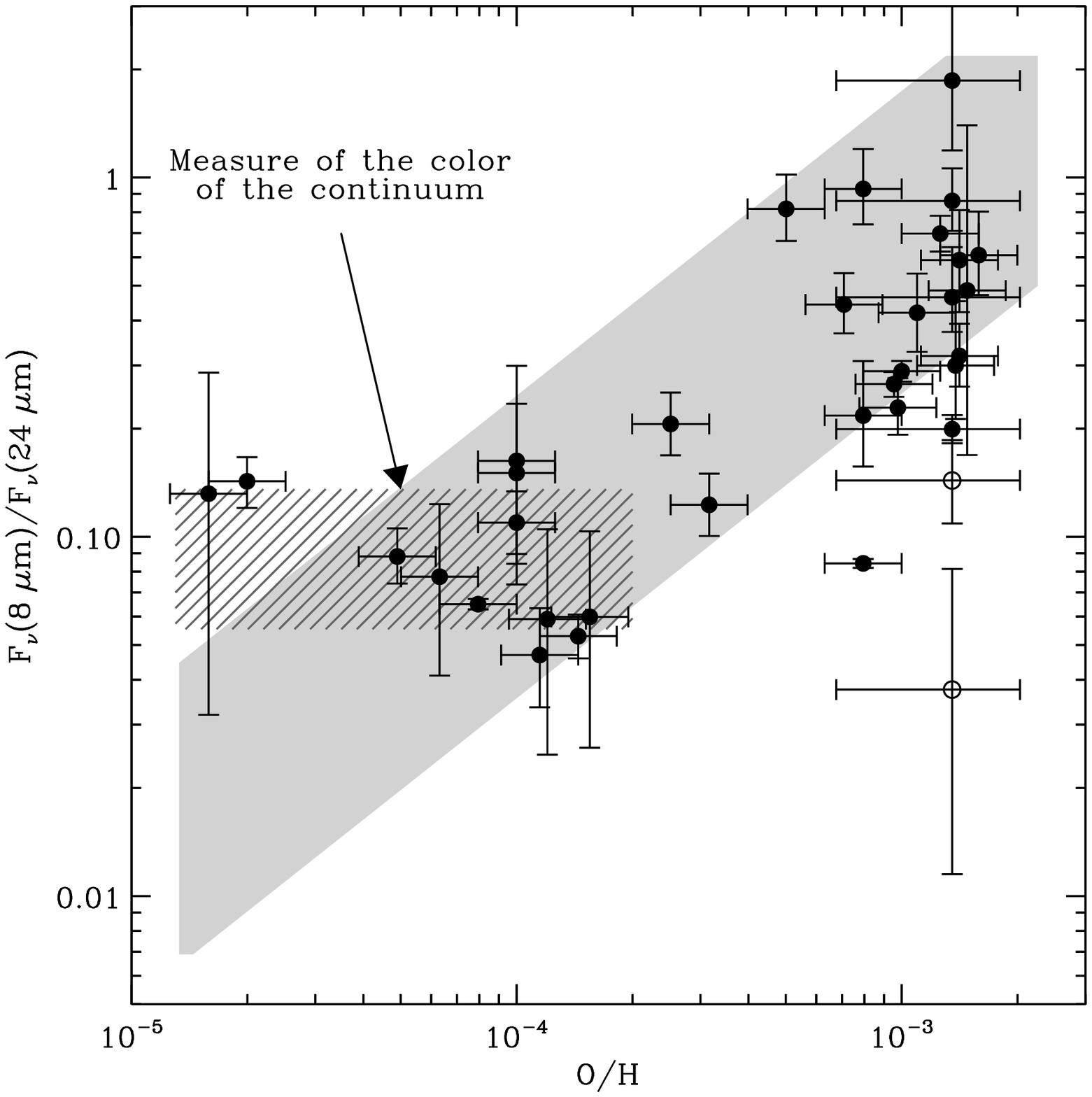}
\includegraphics[width=3.0in]{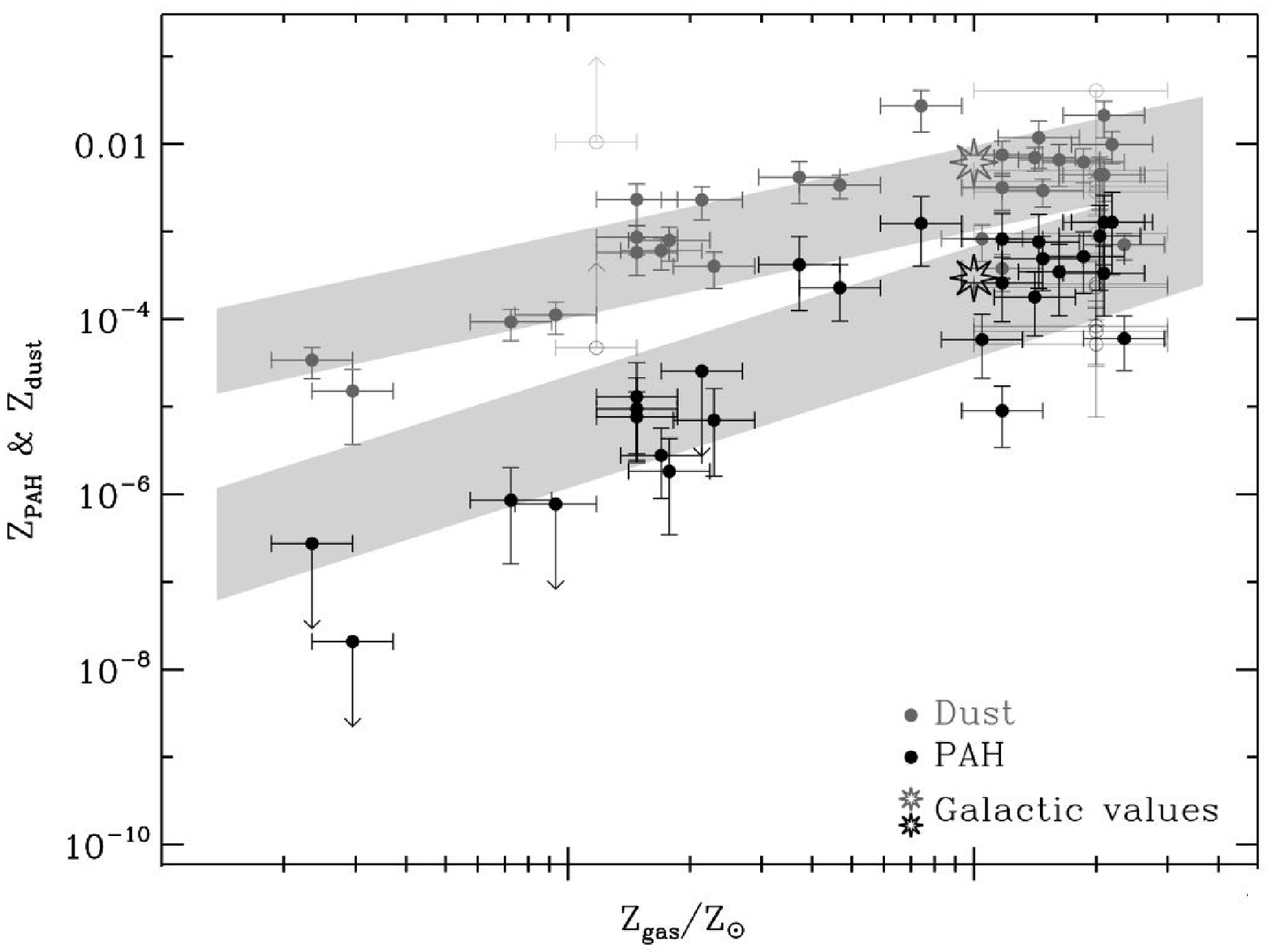}
\end{tabular}
\end{center}
 \caption{{\it Left panel}: The observed correlation between the 8-to-24~\mic\ bands flux density and metallicity. {\it Right panel}: PAH and dust abundances derived from detailed models of galaxies' SED versus galaxies' metallicity.}
    \label{pahvol}
\end{figure}

An exciting discovery made by spectral and photometric observations of nearby galaxies with the {\it Infrared Space Observatory} ({\it ISO}) and {\it Spitzer} satellites was the striking correlation between the strength of their mid-infrared (IR) aromatic features, commonly attributed to the emission from PAHs, and their metallicity, depicted in Figure \ref{pahvol} [left panel; see \cite{galliano08a} for references]. The figure shows the rise of $F_{8/24}$, the 8~\mic-to-24\mic\ band flux ratio with galaxies' metallicity, and the existence of a metallicity threshold below which $F_{8/24}$ is equal the flux ratio of the dust continuum emission. The strength of the aromatic feature is a measure of the PAH abundance. Since PAHs are predominantly made in C-rich AGB stars, this correlation provides the first observational evidence for the delayed injection of AGB condensed dust into the ISM, provided the metallicity is a measure of the galaxies' age. The testing of this hypothesis requires first the determination of the PAH abundance in each galaxy. 

PAHs are very small macromolecules, typically 50~\AA\ in diameter, that are stochastically heated by the ambient radiation field. Consequently, only a fraction of the PAHs are radiating at mid-IR wavelengths at any given time. To determine the total abundance of PAHs, including those too cold to emit in the aromatic features requires the determination of the intensity of the interstellar radiation field (ISRF) to which they are subjected. 

Figure \ref{dustspec} depicts the steps used by \cite{galliano08a} in modeling the galaxies' spectral energy distribution (SED). The galaxy used for this illustrative purpose is the starburst M82. The dust model used in the calculations is the BARE-GR-S model of \cite{zubko04}, consisting of PAHs, and bare silicate and graphite grains with solar abundances constraints. 
  
Figure 7a shows the SED of M82, and its various emission components: stellar emission at optical, and near-IR wavelengths; the PAH spectrum at mid IR wavelengths; dust emission from mid- to far-IR wavelengths; and free-free and synchrotron emission at radio wavelengths. 

Fig 7b depicts a fit to the dust spectrum using an ISRF characterized by a power-law distribution of radiation field intensities. PAH abundance determined by this method will {\it underestimate} the real abundance of PAHs in the galaxy compared to the method outlined below. Our models use the free-free and mid-IR emissions to constrain the gas and dust radiation from the gas and dust from H~II regions, and the far-IR and optical emission to constrain the ISRF that heats the dust in photo-dissociation regions (PDRs). The radio emission is uniquely decomposed into free-free (dashed) and synchrotron (dotted) emission components (see Figure 7d). 

Massive stars are required to produce the ionizing radiation and the expanding SN blast waves that generate, respectively, the observed free-free and synchrotron emission. These stars are produced in an "instantaneous" burst of star formation, in contrast to the stars that are continuously created over the lifetime of the galaxy and mostly contribute to the optical and near-IR emission (Figure 7c).  

The ionizing and a fraction of the non-ionizing radiation emitted by the starburst component that is absorbed in the H~II region is shown as a shaded area in Figure~7d. This energy is reradiated by the dust and gas, giving rise to the thermal IR and free-free emission components shown in the figure. 

The non-ionizing radiation from the older stellar population and the radiation escaping the H~II regions form the diffuse ISRF that is absorbed by the dust in photodissociation regions (PDR) (Figure 7e). The shaded area depicts the fraction of the radiation from the older stellar population that is absorbed in PDRs. The absorbed radiation is reemitted by the dust, giving rise to the IR emission (figure 7f). 

Figure 7g depicts the different emission components, and Figure~7h shows the fit of their sum to the SED of M82. The details of the fitting procedure are described in \cite{galliano08a}.

Using this physical fitting procedure, \cite{galliano08a} derived the abundance of PAHs, silicates, and graphite grains in 35 nearby galaxies with metallicities ranging from 1/50 to 3 times solar. The detailed physical modeling of their SEDs gives larger PAH abundances compared to models that employ a template ISRFs to heat the PAHs and the dust. In these models, such as the one depicted in Fig. 7b, PAHs are subjected to the same intense radiation field as that required to produce the mid-IR emission from hot dust. In our model, PAHs are subjected to a weaker radiation field. Since PAHs do not survive in H~II regions, all their emission originates from PDRs, which are subjected to lower intensity radiation fields than the H~II regions. So, compared to the template ISRF models, a larger amount of PAHs is required to produce the same PAH spectrum with a weaker ISRF. The resulting PAH abundances are plotted versus metallicity in Figure \ref{pahvol} (right panel). The figure shows that the observed trend of increasing 8-to-24~\mic\ band flux ratio with metallicity indeed reflects a trend of increasing PAH abundance with metallicity.  The figure also shows the distinct evolutionary trends of PAHs and the far-IR emitting dust with metallicity.

  \begin{figure}
 \hspace{-0.5in}
  \begin{tabular}{ll}
\includegraphics[width=3.5in]{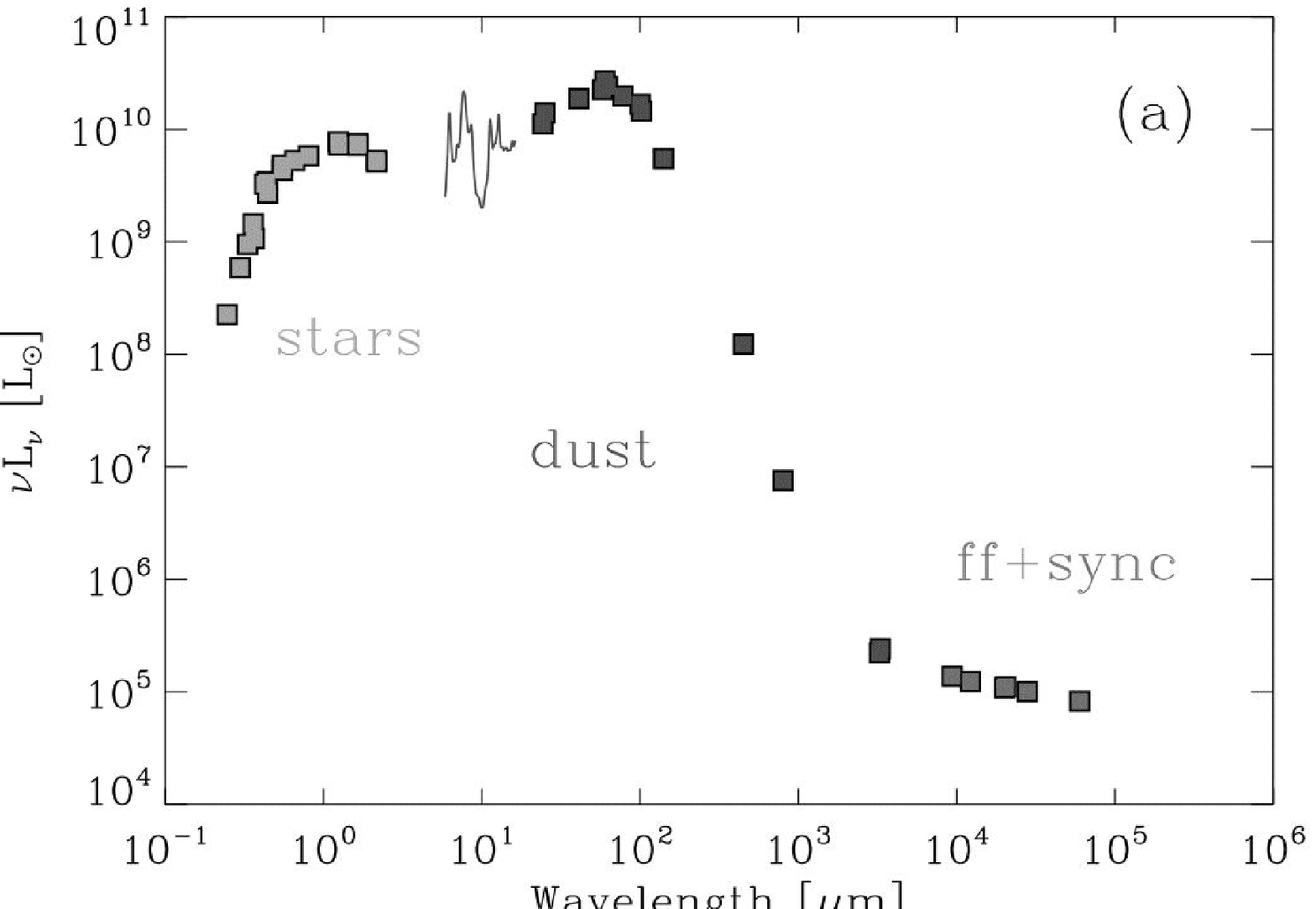}  
\includegraphics[width=3.5in]{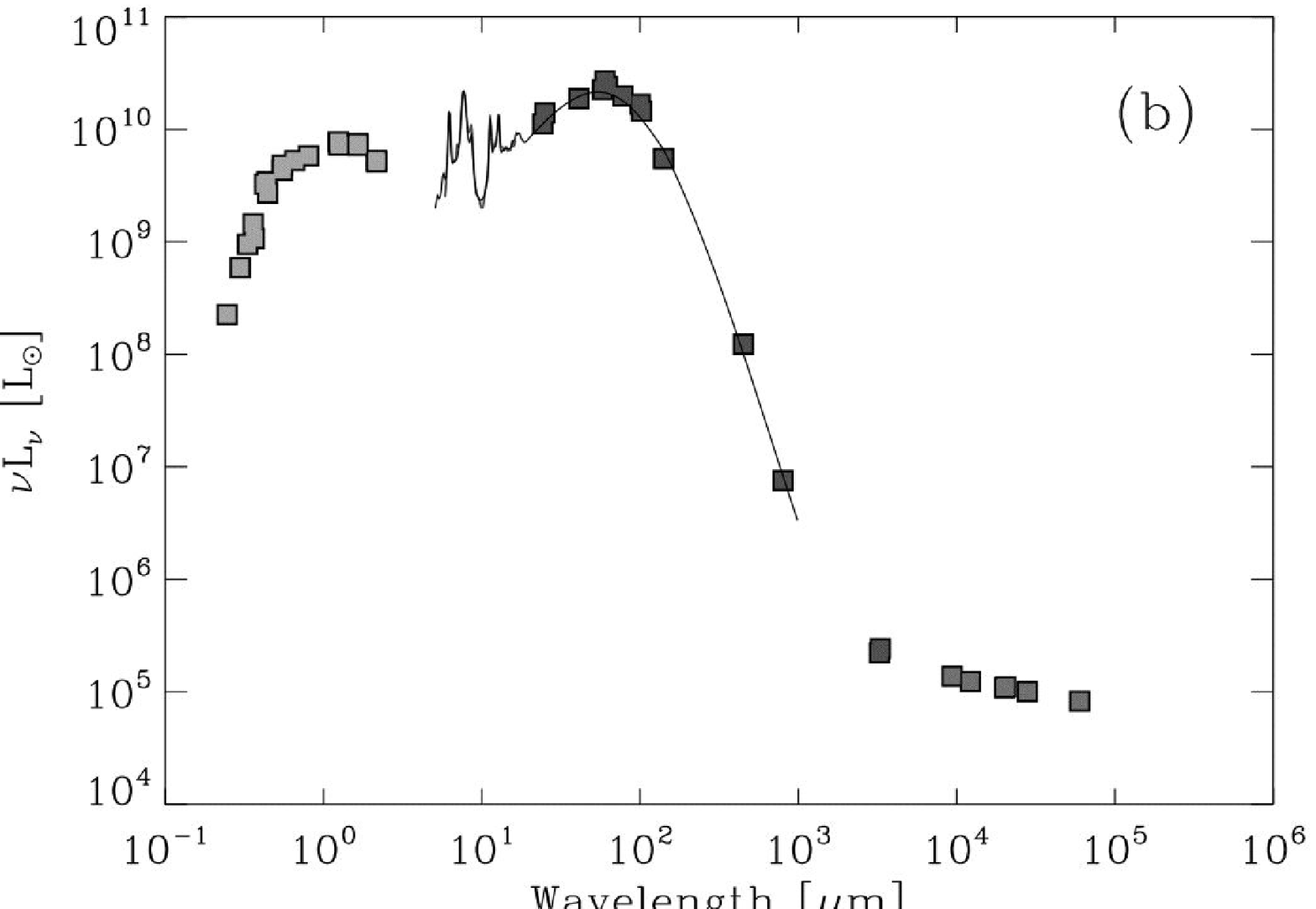} \\ 
\includegraphics[width=3.5in]{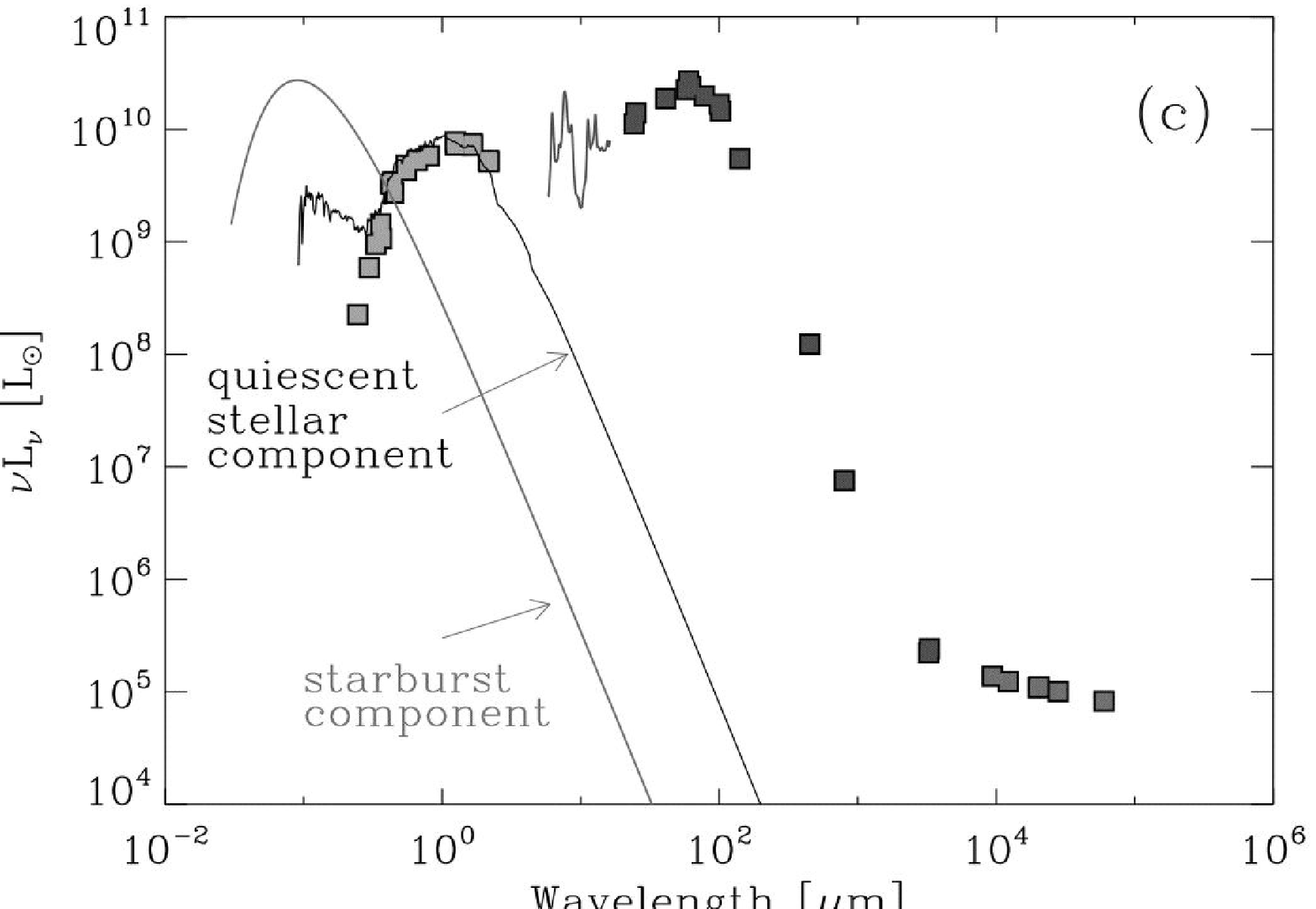}
\includegraphics[width=3.5in]{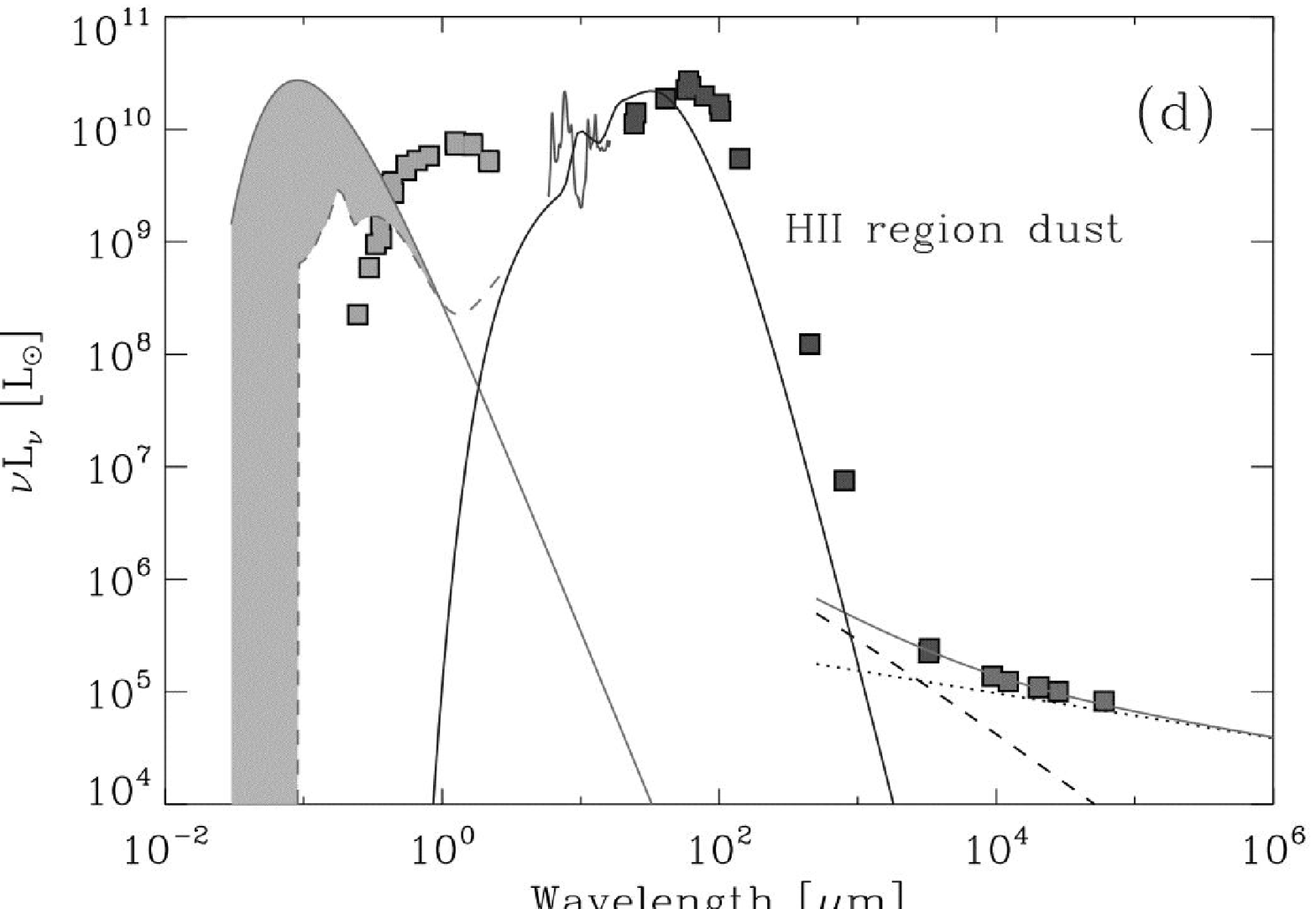} \\ 
\includegraphics[width=3.5in]{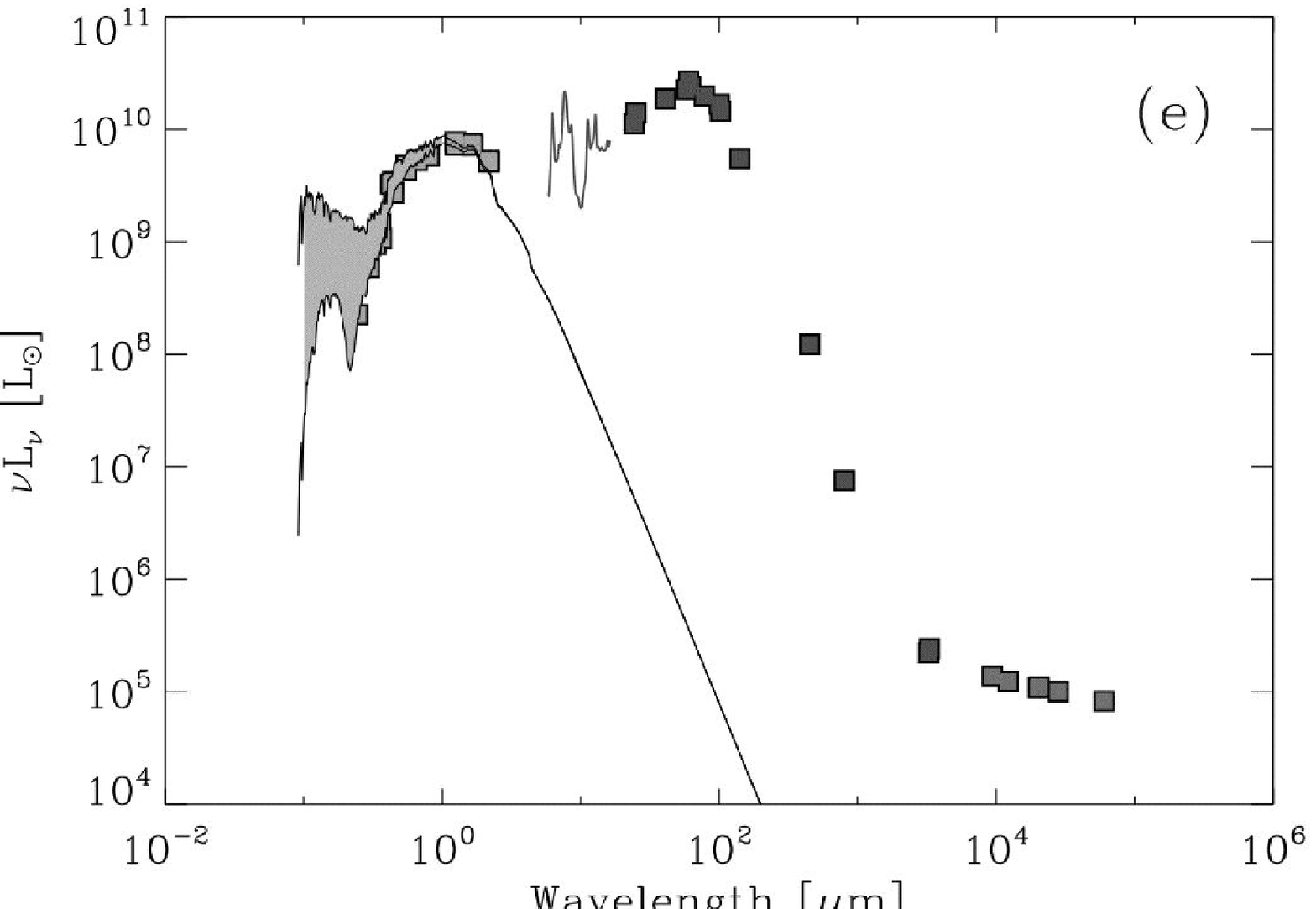}
\includegraphics[width=3.5in]{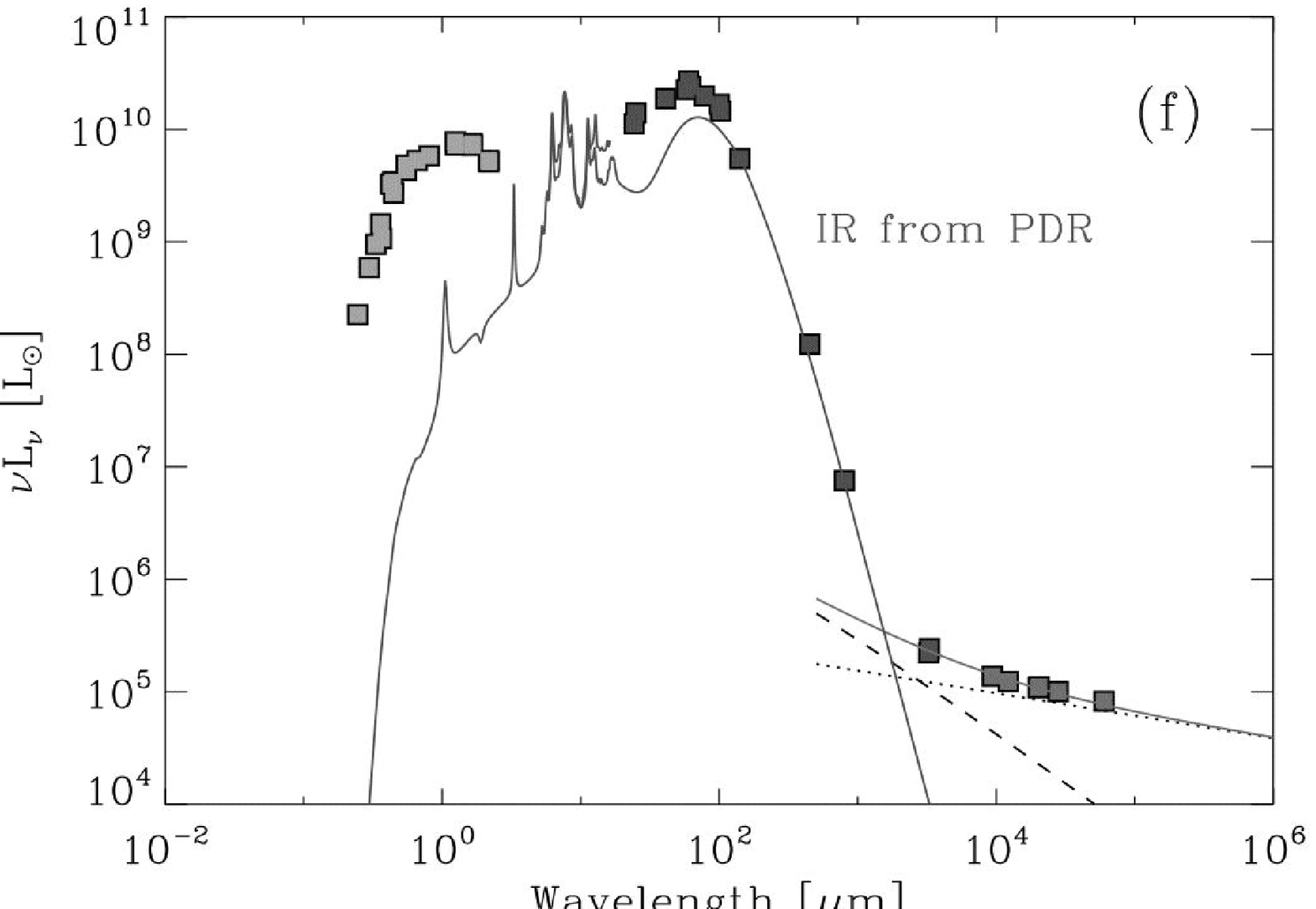} \\ 
\end{tabular}
\end{figure}

  \begin{figure}
  \hspace{-0.5in}
  \begin{tabular}{ll}
\includegraphics[width=3.5in]{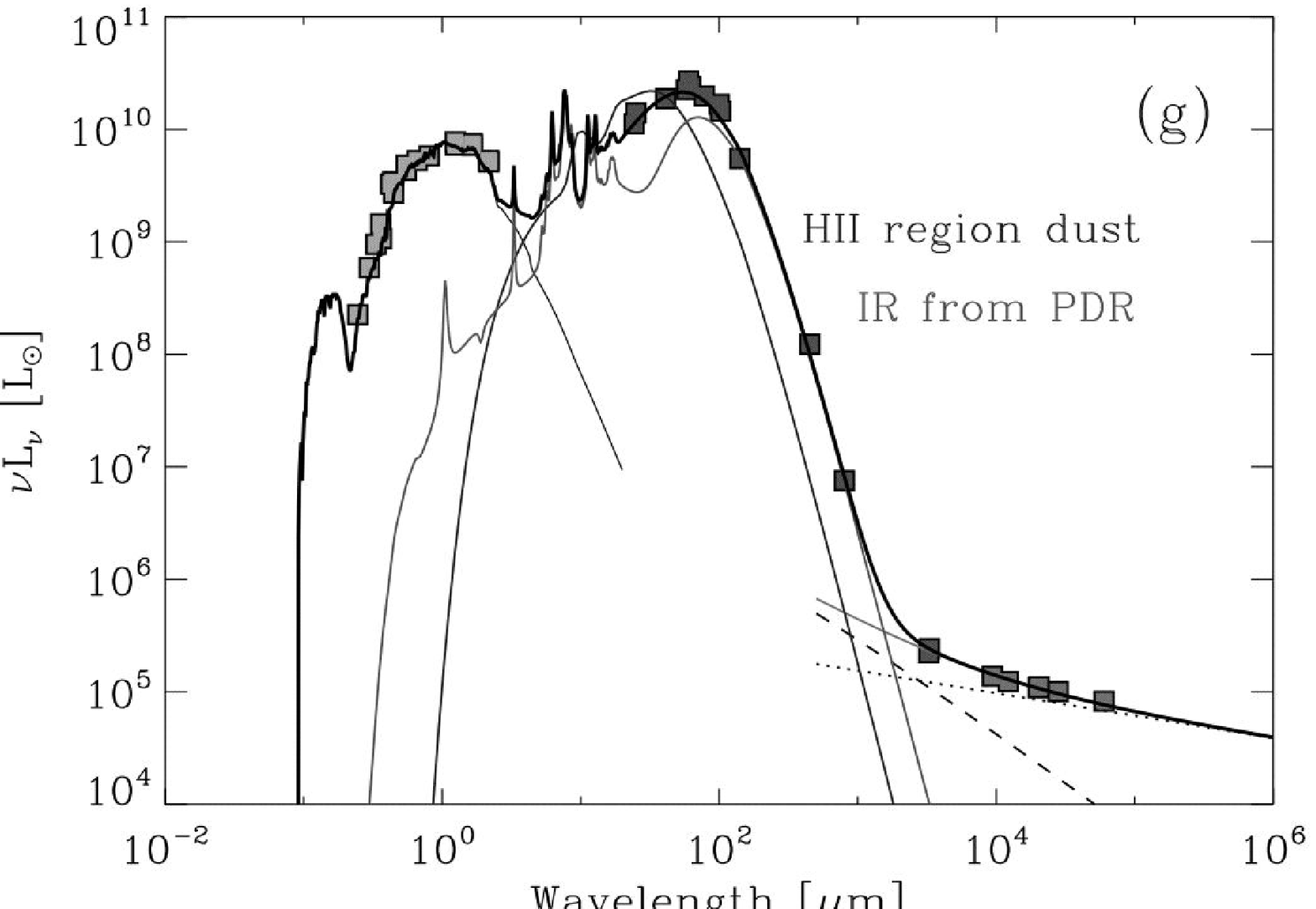}  
\includegraphics[width=3.5in]{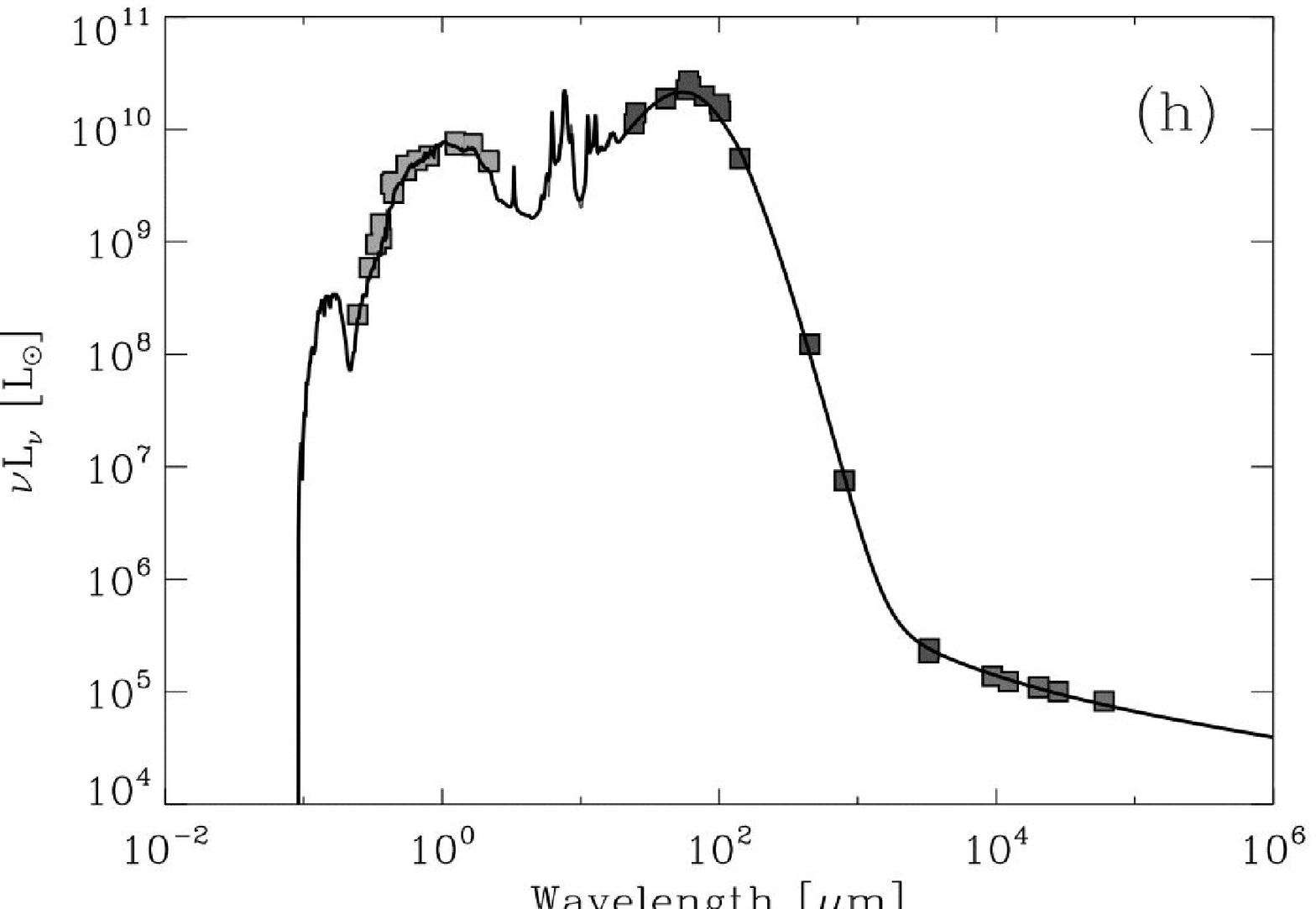}
\end{tabular}
 \caption{ Construction of the fit to the SED of the starburst galaxy M82. See text for details. }
    \label{dustspec}
\end{figure}

\begin{figure}[htbp]
  \begin{center}
\includegraphics[width=5.0in]{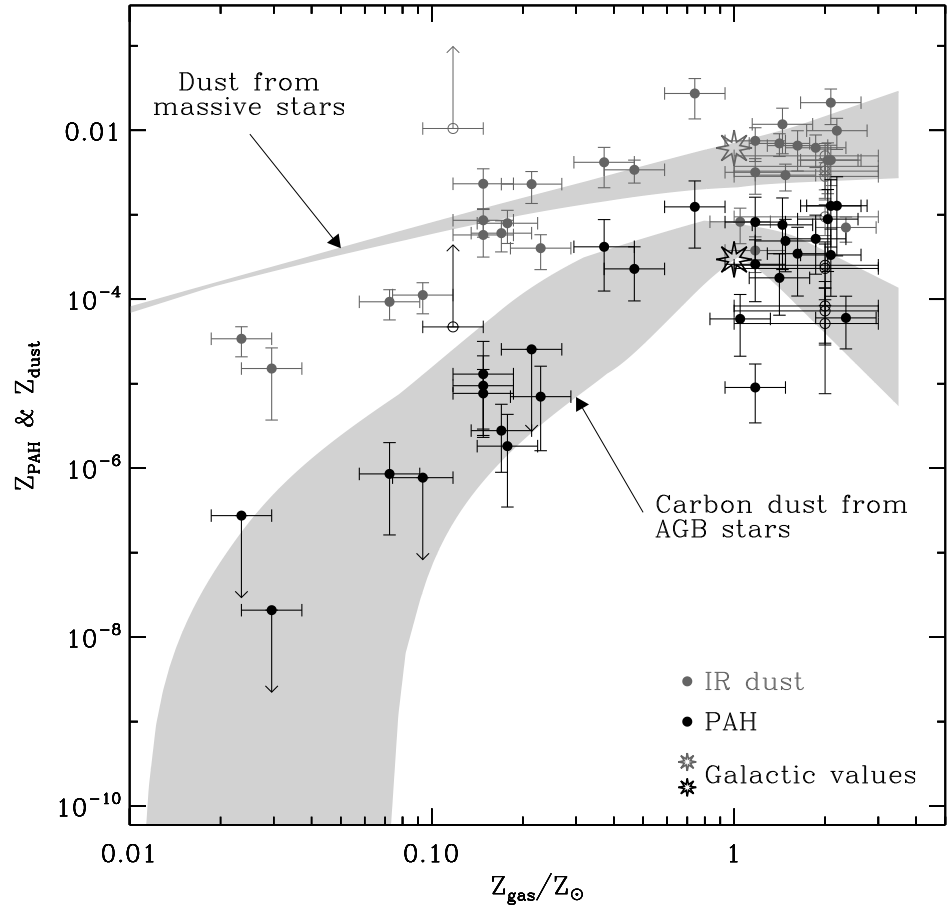}  
\end{center}
 \caption{ Comparison between the metallicity trends of the PAH abundance derived from the observed SED and those derived from the 
           chemical evolution model.
           The shaded area represent the range of model prediction for different grain destruction and star formation rates. Details of the figure are described in \cite{galliano08a}.}
   \label{chemvol}
\end{figure}

Figure \ref{chemvol} compares the evolution of the PAHs and dust components derived from the dust evolution model to the trend of PAH abundances with metallicity. For the sake of this comparison, the evolution of dust abundances as a function of time was converted to an evolution as a function of metallicity, using the age-metallicity relation derived in the model. We emphasize that the parameters used in the dust evolution model (the star formation rate, the stellar IMF) were identical to those used in the population synthesis model that was used to fit the galaxies' SED. From the dust evolution model we already derived the two distinct evolutionary trends of SN- and AGB-condensed dust (see Fig. \ref{pahvol}, right panel for the idealized example). The current figure compares these results with the derived abundance of the PAHs and of the dust that gives rise to the far-IR emission. The latter is dominated by emission from silicates, and should therefore follow the trend of the SN condensates, since most silicate grains are produced in SNe. The figure shows that the observed far-IR emitting dust falls on the evolutionary track of the calculated SN-condensed dust, and that the observed PAH abundances fall on the evolutionary track of the carbon dust that formed in AGB stars. The shaded regions in the figure represent the range of evolutionary tracks that correspond to different parameters that determine the star formation rate and grain destruction efficiencies in the models.   

\section{The presence of massive amounts of dust at high redshift}

The detection of massive amounts of dust in hyperluminous IR galaxies at redshifts $z > 6$ raises challenging questions about the sources capable of producing such large amount of dust during the relatively short lifetime of these galaxies \citep{maiolino06,beelen06,morgan03}. 
For example, the galaxy SDSS J1148+5251 (hereafter \jay) located at $z = 6.4$ was observed at far-IR and submillimeter wavelength \citep{bertoldi03a,robson04,beelen06}. The average IR luminosity of the source is $L_{IR} \sim 2\times 10^{13}$~\lsun, and the average dust mass is $M_d \sim 2\times 10^8$~\msun. Using the \cite{kennicutt98a} relation, one can derive a star formation rate (SFR) of $\sim 3000$~\msun~yr$^{-1}$ from the observed far-IR luminosity. For comparison, the Milky  Way galaxy is about 10~Gyr old, has an average SFR of $\sim 3$~\msun~yr$^{-1}$, and contains about $5\times 10^7$~\msun\ of dust, a significant  fraction of which  was produced in AGB stars.

At $z=6.4$ the universe was only 890~Myr old, using standard $\Lambda$CDM parameters ($\Omega_m = 0.27$, $\Omega_{\Lambda} = 0.73$, and $H_0 = 70$~km~s$^{-1}$~Mpc$^{-1}$). If \jay\ formed at $z = 10$ then the galaxy is only 400~Myr old.
If the SFR had occured at  a constant rate over the lifetime of the galaxy, its initial mass should have been about 10$^{12}$~\msun, which is significantly larger than the dynamical mass $M_{dyn} \approx 5\times 10^{10}$~\msun\ of the galaxy \citep{walter04}. The high observed SFR may therefore represent a recent burst of star formation that has lasted for only about 20~Myr. 
The galaxy \jay\ is therefore at most $\sim 400$~Myr old, and probably significantly younger with an age of only $\sim 20$~Myr. Adopting a current gas mass of $M_g = 3\times 10^{10}$~\msun\ for this galaxy we get that the gas mass fraction at 400~Myr is about 0.60. The dust-to-gas mass ratio is given by $Z_d \equiv M_d/M_g = 0.0067$. 

A significant  fraction of the dust in the Milky Way was produced in AGB stars. However, these stars are not likely  to  contribute significantly to the formation of dust in very young galaxies, since the low mass stars ($M \approx 3$~\msun) that  produce most of the dust did not have time to evolve off the main sequence \citep{dwek98,morgan03,dwek05}.  
In contrast, core collapse SNe ($M > 8$~\msun) and their post-main-sequence progenitors inject their nucleosynthetic products back into the ISM shortly ($t < 20$~Myr) after their formation, resulting in the rapid enrichment of the interstellar medium (ISM) with the dust that formed during the mass loss phase prior to the SN event, or in the explosive SN ejecta. We will hereafter attribute both contribution to the SN event, since both are return "promptly" to the ISM. But can SNe account for the large amount of dust seen in this object? The answer to this question is complicated by the fact that SNe are also the main source of grain destruction during the remnant phase of their evolution \citep{jones96,jones04}. The problem can therefore only be quantitatively addressed with CE models for the dust in these systems.

The results of the detailed dust evolution models described summarized in Figure \ref{chemvol} show that the contribution of AGB stars to metal and dust abundance can  be neglected in galaxies with ages less than about 400~Myr. 
The equations for the chemical evolution of the galaxy can then be considerably simplified using the instantaneous recycling approximation, which assumes that stars return their ejecta back to the ISM promply after their formation. The evolution of the dust abundance can then be written in analytical form \citep{morgan03,dwek07b}.

In particular,  the yield of dust, \ydust, required to obtain a given dust-to-gas mass ratio, $Z_d$,  when the galaxy reaches a given gas mass fraction \mug, is given by \citep{dwek07b}:
\begin{equation}
\yduste =Z_d\ \left[{\misme + R\ \mstare \over 1-\muge^{\nu-1}}\right]
\label{ydzd}
\end{equation}

\noindent where,
\begin{equation}
\nu \equiv {\misme + \mstare \over (1-R)\ \mstare}\qquad ,
\label{nu_eq}
\end{equation}
$R$ is the fraction of the stellar mass that is returned back to the ISM during the stellar lifetime, \mism\ is given by eq. (2), and \mstar\ is the mass of all stars born per SN event. For example, $\mstare=147$ and $50$ ~\msun, respectively, for a Salpeter and top-heavy IMF.

\begin{figure}
    \begin{center}
\includegraphics[width=5.0in]{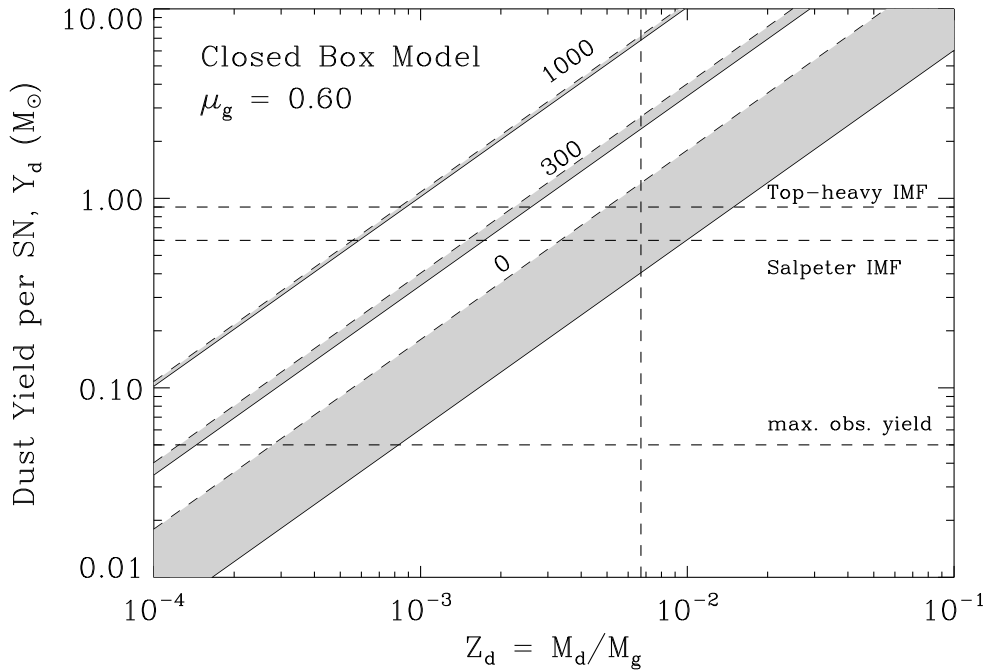} 
  \end{center} 
 \caption{The IMF-averaged yield of dust by type~II supernova, \yd, that is required to account for a given dust-to-gas mass ratio $Z_d$, is presented for different values of \mism\ given in units of \msun. Solid and dashed lines correspond to calculations done for a top-heavy and a Salpeter IMF, respectively. The horizontal dashed line near the bottom of the figure corresponds a value of $Y_d = 0.054$~\msun, the highest inferred yield of dust in supernova ejecta to date \cite{rho08}. The vertical dotted line represents the value of $Z_d$ at $\muge =0.60$. Curves are labeled by \mism\ given  in units of \msun.  The top two dashed (solid) horizontal lines represent IMF-averaged theoretical dust yields for a Salpeter (top-heavy) IMF, assuming 100\% condensation efficiency in the SN ejecta. }
   \label{snyield}
\end{figure}

Figure \ref{snyield} shows how much dust an average SN {\it must} produce in order to give rise to a given dust-to-gas mass ratio, for various grain destruction efficiencies. The value of \ydust\ was calculated when \mug\ reaches a value of 0.60, the adopted gas mass fraction of \jay\ at 400~Myr. 
Calculations were performed for two different functional forms of the stellar IMF: a Salpeter IMF in which $\phi(m) \sim m^{-2.35}$ and $0.1 < m$(\msun) $< 100$; and a top heavy IMF characterized by the same mass limits but a flatter slope $\phi(m) \sim m^{-1.50}$. Here, $\phi(m)$ is the number of stars per unit mass interval, normalized to unity between 0.1 and 100~\msun.

The figure shows that, for example, to produce a value of $Z_d = 0.0067$ at \mug\~=~0.60, a SN must produce about 0.4~(1.2)~\msun\ of dust for a top-heavy (Salpeter) IMF, provided the dust is not destroyed in the ISM, that is, \mism\ = 0. Even with modest amount of grain destruction, \mism\ = 100~\msun, the required SN dust yield is dramatically increased to about $1-2$~\msun, depending on the IMF. 
The horizontal line in the figure corresponds to a value of \ydust\ = 0.054~\msun, the largest mass of SN-condensed dust inferred to be present in a supernova or SNR \citep{rho08,sugerman06}. Contrary to the claim by \cite{rho08}, this yield is not sufficient to account for the large amount of dust observed in high redshift galaxies, since the quoted chemical evolution models of \cite{morgan03} do not include the effect of grain destruction. The figure shows that even without grain destruction, the largest observed yield can only give rise to a dust-to-gas mass ratio of $\sim 4\times10^{-4}$. If the mass of dust in the ejecta of Cas~A represents a typical SN yield, then other processes, such as accretion onto preexisting grains in molecular clouds is needed to produce the mass of dust in J1148+5251.

\section{Summary}

Dust evolution models have proven to be very successful in predicting global evolutionary trends in dust abundance and composition, and in analyzing the origin of dust in the early universe. 
An important prediction of these models is that SN- and AGB-condensed dust should follow distinct evolutionary paths because of the different stellar evolutionary tracks of their progenitor stars. By analyzing the UV-to-radio SED of 35 nearby galaxies we have identified silicates and PAHs, respectively, as tracers of SN- and AGB-condensed dust. Our SED fitting procedure used chemical evolution, dust evolution, and population synthesis models in a consistent fashion. The models used the free-free and mid-IR emissions to constrain the gas and dust radiation from the gas and dust from H~II regions, and the far-IR and optical
emission to constrain the ISRF that heats the dust in PDRs.
The observed correlation of the intensity of the mid-IR emission from PAHs with their metallicity can then be interpreted as the result of stellar evolutionary effects which cause the delayed injection of carbon dust into the interstellar medium.

The early universe is a unique environment for studying the role of massive stars in the formation and destruction of dust. 
 The equations describing their chemical evolution can be greatly simplified by using the instantaneous recycling approximation, and by neglecting the delayed contribution of low mass stars to the metal and dust abundance of the ISM. Neglecting any accretion of metals onto pre-existing dust in the interstellar medium, the evolution of the dust is then primarily determined by the condensation efficiency of refractory elements in the ejecta of Type~II supernovae, and the destruction efficiency of dust by SN blast waves.

We applied our general results to \jay, a dusty, hyperluminous quasar at redshift $z = 6.4$ and found that the formation of a dust mass fraction of $Z_d = 0.0067$ in a galaxy with an ISM mass of $3\times 10^{10}$~\msun, requires an average SN to produce between 0.5 and 1~\msun\ of dust if there was no grain destruction. Such large amount of dust can be produced if if the condensation efficiency in SNe is about unity. Observationally, the required dust yield is in excess of the largest amount of dust ($\sim 0.054$~\msun) observed so far to have formed in a SN. This suggests that accretion in the ISM may play an important role in the growth of dust mass.
For this process to be effective, SNRs must significantly increase, presumably by non-evaporative grain-grain collisions during the late stages of their evolution, the number of nucleation centers onto which refractory elements can condense in molecular clouds.

\acknowledgments
This work was supported by NASA's LTSA 03-0000-065. 

\newpage




\end{document}